\documentclass[prl,twocolumn,showpacs,preprintnumbers,amsmath,amssymb]{revtex4}
\usepackage{graphicx}
\usepackage{epsfig}
\begin{document}
\title{Time-dependent Ginzburg-Landau theory with floating nucleation kernel;\\
FIR conductivity in the Abrikosov vortex lattice}
\author{Pei-Jen Lin$^1$}
\author{P. Lipavsk\'y$^{2,3}$}
\address{$^1$NCTS, National Tsing Hua University,
Hsinchu 300, Taiwan\\
$^2$ Faculty of Mathematics and Physics, Charles University, 
Ke Karlovu 3, 12116 Prague 2, Czech Republic\\
$^3$Institute of Physics, Academy of Sciences, 
Cukrovarnick\'a 10, 16253 Prague 6, Czech Republic}

\begin{abstract}
We formulate the time-dependent Ginzburg-Landau theory, with 
the assumption of local equilibrium made in the reference
frame floating with normal electrons. This theory with 
floating nucleation kernel
is applied to the far infrared (FIR) conductivity in 
the Abrikosov vortex lattice. It yields better agreement with
recent experimental data [PRB {\bf 79}, 174525 (2009)] than the
customary time-dependent Ginzburg-Landau theory.
\end{abstract}
\maketitle

The time-dependent Ginzburg-Landau (TDGL) equation is a 
useful extension of the equilibrium Ginzburg-Landau theory. 
Unfortunately, microscopic derivations 
\cite{Sch66,AT66,GE68,MRE93,HYY09} 
guarantee its validity under such restrictive conditions that it 
seems more difficult to find justified nontrivial applications 
than to solve it. The {\mbox TDGL} equation is thus most often
applied beyond its nominal range of validity. 

As one leaves the familiar vicinity of the 
superconducting phase transition and asymptotically slow
processes, the intuitive foundation of the theory becomes
shaky.  The TDGL theory contains an assumption of local equilibrium,
which is dependent on reference frame;
when we adapt the equilibrium-based equation to 
non-equilibrium problems, we should at least work in 
the reference frame in which electrons are as close 
to local equilibrium as possible. This is the frame 
floating with the normal current
in the background of a superconducting condensate. 
To this end, in this paper we introduce what we 
refer to as a {\em floating nucleation kernel}.

The standard TDGL theory is formulated using a kernel static 
in the laboratory system. We will show that compared to the TDGL 
theory in the floating system, the laboratory formulation lacks 
a term which is particularly important at high frequencies 
of the driving field. We will demonstrate the effects of this term  on 
the conductivity in the sub-gap far-infrared (FIR) region. 
Comparing our results with recent FIR magneto-transmission 
measurements of Ikebe {\em et al} \cite{ISIFK09}, we will show 
that use of the floating nucleation kernel improves agreement 
between the theory and experimental data. 

Let us first describe the magneto-transmission measurement.
It is performed on a thin layer perpendicularly penetrated 
by the magnetic field in the form of vortices. The incident 
FIR light is perpendicular to the surface and its electric field
drives currents which determine the amplitude and phase of
the transmitted light which is measured.

Both the normal and the superconducting electrons are 
accelerated by the electric field and experience 
a friction with the lattice. The friction of the condensate 
is much weaker since Joule heat develops only in 
vortex cores moving perpendicularly to the electric field. 
The relative contribution of these components to the current 
depends on the frequency of the driving field; the higher the
frequency the higher will be the fraction of the normal current.

It is useful to inspect characteristic times for NbN, the 
material used by Ikebe {\em et al} \cite{ISIFK09}. 
The optical gap $2\Delta=5.3$~meV implies the maximal sub-gap 
frequency $\omega< 10$~THz. The mean time between two 
collisions of the normal electron is $\tau_n\sim 5$~fs, 
therefore during a \mbox{single} period of the sub-gap FIR field the 
electron  loses momentum more than a hundred times. At zero 
magnetic field the condensate suffers no friction. The field of 
amplitude $E$ accelerates the condensate to velocity 
$e^*E/\omega m^*$, while a normal electron is accelerated 
to $eE\tau_n/m$. At the  measurement  temperature, $T=3$~K and $T_c=15$~K, 
the density of condensed electrons exceeds the
normal density, therefore the condensate clearly dominates the
total current. A different situation obtains, however, for the 
Joule heat. The condensate current is out of phase with the 
driving electric field and generates no heat. The normal current
is in-phase, producing heat. If the magnetic field penetrates
the sample, the condensate generates the Joule heat due to motion 
of vortices. We will see that for the sub-gap FIR frequencies 
the Joule heat value is much smaller than the amount of heat 
generated by normal electrons. 

To identify the Joule heat, it is necessary to measure 
the transmission coefficient, including its phase. This 
allows one to determine the complex conductivity $\sigma$ 
with ${\rm Im}\,\sigma$ giving the off-phase current and 
${\rm Re}\,\sigma$ 
for the in-phase current. Ikebe {\em et al} \cite{ISIFK09} 
achieved this task by splitting short pulses and mixing 
them again after one of branches passed through the sample. 
As mentioned, we will compare their experimentally established 
$\sigma$ with theoretical predictions based on the TDGL theory 
in the laboratory and the floating coordinate system.

We will use the electric field ${\bf E}(\tau)={\rm Re}\, 
\left[{\bf E}{\rm e}^{-i\omega\tau}\right]$ and current
${\bf J}(\tau)={\rm Re}\,\left[{\bf J}{\rm e}^{-i\omega\tau}
\right]$. 
The complex conductivity is defined
via ${\bf J}=\sigma\,{\bf E}$. The current has a small Hall 
component which we neglect in our discussion for convenience. 

The TDGL equation derived using the static kernel 
\cite{Tinkham},
\begin{align}
&{1\over 2m^*}\left(\!-i\hbar\nabla\!-\!
\frac{e^*}{c}{\bf A}\right)^2\psi +\alpha\psi+
\beta\left|\psi\right|^2\psi=-
\Gamma\partial_\tau\psi,
\label{lin1}
\end{align}
describes the evolution of the condensate including a relaxation 
of the GL function $\psi$ towards its equilibrium value. The
vector potential is that of the internal magnetic field as well
as the electric field of the FIR light ; ${\bf B}=\nabla\times{\bf A}$
and ${\bf E}=-(1/c)\partial_\tau{\bf A}$. The electric current 
\begin{align}
{\bf j}_s={e^*\over m^*}{\rm Re}~\left[ \bar\psi\left(-i\hbar\nabla-
\frac{e^*}{c}{\bf A}\right)\psi \right]
\label{lin2}
\end{align}
is composed of circulating diamagnetic currents and oscillating 
response to the light. We solve Eq.~\eqref{lin1} to linear
order in $\bf E$ and eliminate the diamagnetic currents by
averaging over the elementary cell of the Abrikosov vortex 
lattice; ${\bf J}_s=\left\langle{\bf j}_s\right\rangle=
(B/\Phi_0)\int_{\rm cell}{\rm d}x{\rm d}y\,{\bf j}_s$.
The supercurrent, ${\bf J}_s=\sigma_s{\bf E}$, gives the 
condensate conductivity
\begin{equation}
\sigma_s= \frac{3\sigma_0}{\beta_{\rm A}}
\frac{1-t-b}{b-i\omega \tau_s},
\label{lin3}
\end{equation}
where $t=T/T_c$, $b=B/H_{c2}$ are the dimensionless temperature 
and magnetic field, $\sigma_0$ is the normal state conductivity, 
$\beta_{\rm A}=1.16$ is the Abrikosov constant for hexagonal 
vortex lattice, and $\tau_s=\Gamma (1-t)/\alpha$. Deriving 
Eq.~\eqref{lin3} we have used the GL parameter \cite{BR05}
\begin{equation}
\Gamma={12\pi\sigma_0\alpha\kappa^2\xi^2\over c^2(1-t)^2}.
\label{lin4}
\end{equation}
The zero-temperature coherence length is determined by the
upper critical field; $\xi^2=\Phi_0/(2\pi H_{c2}^0)$. 
Here $H_{c2}^0=15$~T is obtained via the linear extrapolation 
$H_{c2}=H_{c2}^0(1-t)$ from experimental data in Fig~3 of
\cite{ISIFK09}. The normal-state conductivity 
$\sigma_0=2\cdot 10^4/\Omega$cm, experimentally established 
at 20~K \cite{ISIFK09}, has weak temperature dependence 
and can be used at 3~K.

  \begin{figure}[h]  
  \centerline{
  \psfig{figure=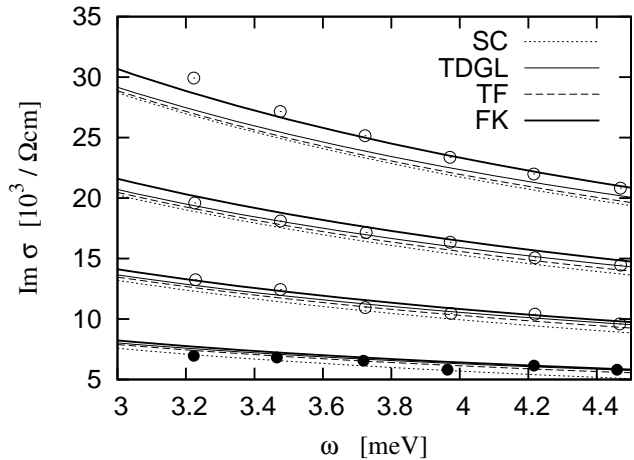,width=6.3cm,angle=-90}}
\vskip 2pt
\caption{Imaginary part of the conductivity giving 
non-dissipative currents: Thin lines are the superconducting condensate
conductivity ${\rm Im}\,\sigma_s$ (dotted), the TDGL conductivity
${\rm Im}\,\sigma_{\rm GL}$ (full), and the two-fluid modification
of the TDGL conductivity ${\rm Im}\,\sigma_{\rm tf}$ (dashed). 
The heavy line is the conductivity 
${\rm Im}\,\sigma_{\rm fk}$ evaluated in the floating system. 
Experimental data of Ikebe {\em et al} \cite{ISIFK09} at 
7~T ($\bullet $) are in the nominal validity range of the 
TDGL theory, while the lower magnetic fields 5~T, 3~T, 
and 1~T ($\circ $) are not.}
\label{fig1}
\end{figure}

In Fig.~\ref{fig1} one can see that the imaginary part 
of $\sigma_{\rm s}$ from formula \eqref{lin3} reproduces 
recent experimental data of Ikebe {\em et al} \cite{ISIFK09}. 
Here we use the GL parameter $\kappa=40$, the only fitting 
parameter in the present theory. It is adjusted to fit the 
imaginary part of the conductivity 
at 7~T. Our main interest is in the Joule heat given by the 
real part of the conductivity. 

Formula \eqref{lin3} was derived for the dense Abrikosov 
vortex lattice.
Theoretically, the region of nominal validity is
$B>4$~T, at the temperature $T=3$~K. It is therefore somewhat
surprising that theoretical curves of ${\rm Im}\,\sigma$ 
slightly depart from the experimental data only at the 
lowest magnetic field $B=1$~T.

\begin{figure}[h]  
  \centerline{
  \parbox[t]{9cm}{
  \psfig{figure=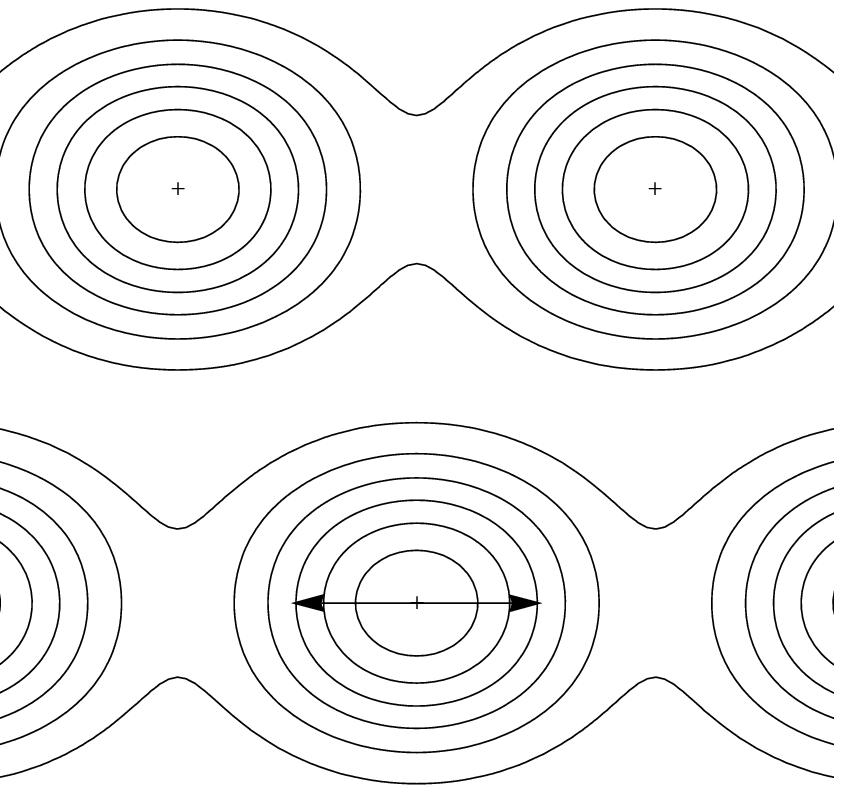,width=4cm}
  \psfig{figure=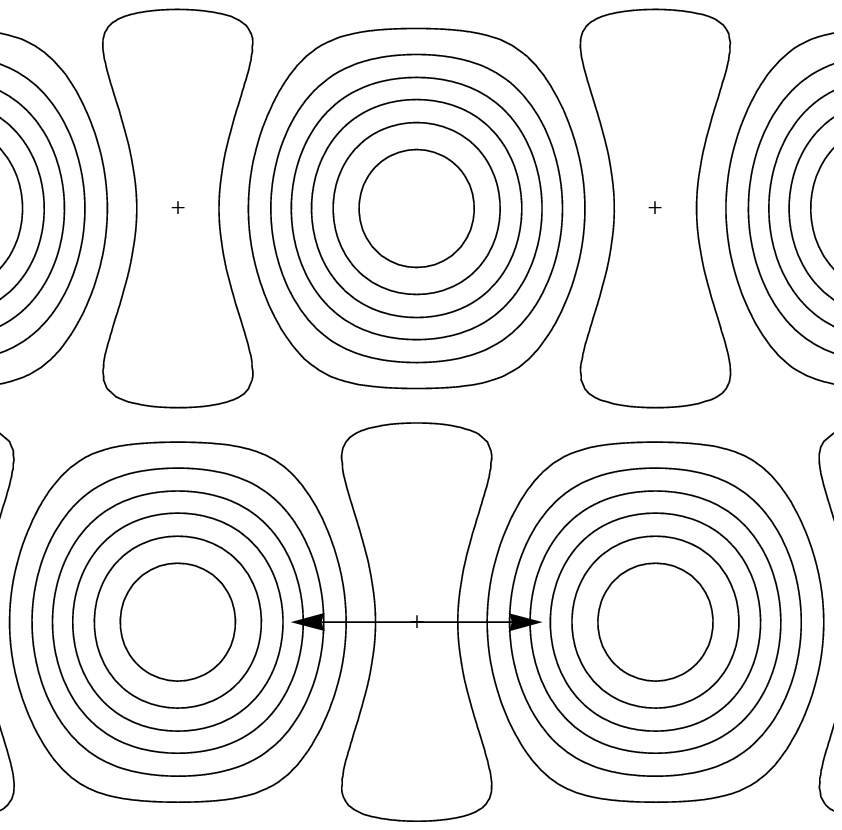,width=4cm}}}
  \vskip 2pt
\caption{Heat production (left) and the power absorption 
(right) in the hexagonal Abrikosov vortex lattice: Crosses 
denote centers of vortices. The electric field is 
polarised vertically so that vortices oscillate horizontally
with amplitude shown by arrows. The Joule heat is produced 
at vortex cores, their horizontal motion is responsible for
elongation of the heated region. Absorption of power is rather 
delocalised. Its maxima are also around vortex cores but 
elongated vertically. The rounded minima are between vortices. 
Difference of these two maps shows that the `rigid' GL function 
transfers the power to be dissipated in cores.
}
\label{fig2}
\end{figure}

Due to the relaxation term $\Gamma\partial_t\psi$, the TDGL
equation \eqref{lin1} includes a damping and generates 
Joule heat \cite{KeS98}, $\dot Q=4k_{\rm B}T\Gamma(\omega/2\pi)
\left\langle|\partial_\tau\psi|^2\right\rangle$, where the 
brackets denote the time average: $\langle\phi\rangle\equiv
(\omega/2\pi)\int_0^{2\pi/\omega}{\rm d}\tau \phi$. The left-hand 
panel of 
Fig.~\ref{fig2} shows that the supercurrent produces Joule 
heat only at vortex cores. The right-hand panel of Fig.~\ref{fig2} 
presents the spatial distribution of the power 
absorbed by the condensate from the electric field $W=
\left\langle{\bf j}_s\cdot{\bf E}\right\rangle$. 
The most intensive 
absorption is around vortices in regions elongated in the
vertical direction which is parallel to the electric field. 
Deep minima of the absorption are between vortices in 
horizontal rows. Comparing the two panels 
shows that the relation between absorption and heat 
production is very non-local.

\begin{figure}[h]  
  \centerline{
  \psfig{figure=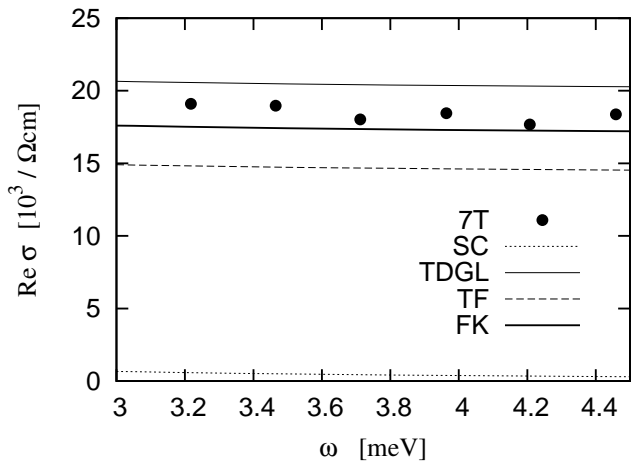,width=6.3cm,angle=-90}}
\vskip 2pt
\caption{Real part of the conductivity giving  Joule heat:
Points are experimental data of Ikebe {\em et al}
\cite{ISIFK09} for 7~T ({$\bullet $}). The superconducting 
condensate contribution (dotted line) given by formula \eqref{lin3} 
is by an order of magnitude too small. The time-dependent 
Ginzburg-Landau theory (thin line) adds a contribution of normal 
electrons, see Eq.~\eqref{lin5}, arriving at too high values. The 
two-fluid approach (dashed line) reduces the conductivity subtracting
double-counted condensed electrons from the normal conductivity,
see Eq.~\eqref{lin7}. The floating kernel approach (heavy line) given 
by Eq.~\eqref{lin11} removes double-counting
from the supercurrents and yields the closest agreement with
experiment.
}%
\label{fig3}
\end{figure}

The fraction of Joule heat due to the condensate is small. 
In Fig.~\ref{fig3} we compare the real part of the condensate 
conductivity \eqref{lin3} with experiment. Indeed, the 
discrepancy between experimental data and Re\,$\sigma_s$ 
indicates that the supercurrent produces only a minor part of 
the Joule heat; the normal current  cannot be neglected .

\begin{figure}[h]  
  \centerline{
  \psfig{figure=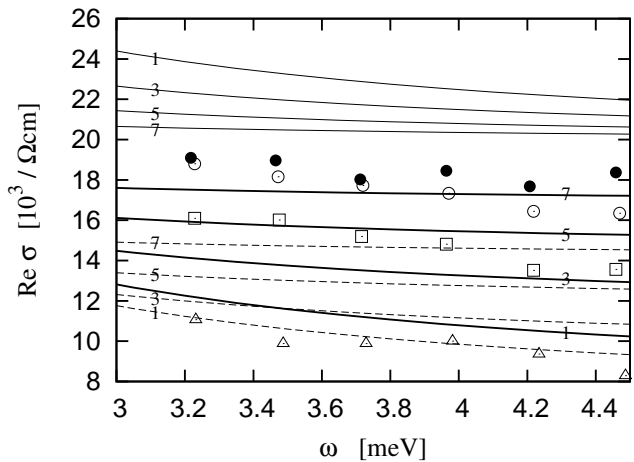,width=6.3cm,angle=-90}}
\vskip 2pt
\caption{Real part of the conductivity giving the Joule heat:
Points are experimental data of Ikebe {\em et al}
\cite{ISIFK09} for 7~T ({$\bullet $}), 5~T ({$\circ $}), 
3~T ($\Box $), and 1~T ($\triangle $). The time-dependent Ginzburg-Landau
theory (thin line) given by \mbox{Eq.}~\eqref{lin5} overestimates the 
dissipation. The two-fluid approach given by Eq.~\eqref{lin7} reduces 
the dissipation too much leading to the underestimate. 
The floating kernel approach ( heavy  line) given by 
\mbox{Eq.}~\eqref{lin11} yields higher values although still smaller
than experimental data.}%
\label{fig4}
\end{figure}

From microscopic derivations \cite{Sch66,AT66,GE68,Kop01} 
of the GL theory it follows that the normal current and the 
supercurrent simply add. Adding the current ${\bf J}_n=\sigma_0
(1+i\tau_n\omega){\bf E}$ which would appear in the normal state 
one obtains the TDGL conductivity 
\begin{equation}
\sigma_{\rm GL}=\sigma_s+\sigma_n,
\label{lin5}
\end{equation}
with the normal conductivity $\sigma_n=\sigma_0(1+i\tau_n\omega)$. 
For experimentally established values $\sigma_0=2\cdot 10^4/\Omega$cm 
and $\tau_n=5$~fs \cite{ISIFK09}, the normal conductivity yields 
a negligible contribution to ${\rm Im}\,\sigma_{\rm GL}$,  as seen in 
Fig.~\ref{fig1}, but it provides the dominant contribution to 
${\rm Re}\,\sigma_{\rm GL}$. One can see in Fig.~\ref{fig3} that 
${\rm Re}\,\sigma_{\rm GL}$ is much closer to observed values than 
${\rm Re}\,\sigma_s$. It is higher than the observed values, 
however. This problem becomes more serious at lower magnetic
fields, where the observed real part of total conductivity is 
further reduced well below the level of the normal conductivity, see 
Fig.~\ref{fig4}, while the TDGL conductivity is always larger,
${\rm Im}\,\sigma_{\rm GL}>{\rm Im}\,\sigma_n$.

The simple addition of normal  current  and supercurrent
works well close to the phase transition but it badly 
overestimates conductivity far from it. 
Apparently, it is insufficient simply to add the supercurrent 
and the
normal current; the electric field accelerates all electrons.
Since electrons in the condensate escape frictional effects,
this fraction of electrons must be removed in order to obtain
the normal conductivity. An intuitive way to avoid 
double-counting of condensed electrons is to introduce a 
normal current reduced in the spirit of the two-fluid
model,
\begin{equation}
\tilde{\bf j}_n=\left(1-{2|\psi|^2\over n}\right){\bf J}_n.
\label{lin6}
\end{equation}
The total current averaged over the elementary vortex lattice cell, 
${\bf J}={\bf J}_s+\tilde{\bf J}_n$, leads to a conductivity 
\begin{equation}
\sigma_{\rm tf}=\sigma_s+\left(t+b\right)\sigma_n,
\label{lin7}
\end{equation}
where we have evaluated the averaged normal fraction, 
$1-2\left\langle|\psi|^2\right\rangle/n=t+b$. 
One can see in Figs.~\ref{fig1} and~\ref{fig3} that the two-fluid 
conductivity yields the same non-dissipative currents described by 
${\rm Im}\,\sigma_{\rm tf}$ as the TDGL theory, but that it allows for
${\rm Re}\,\sigma_{\rm tf}$ smaller than the normal conductivity. In
fact ${\rm Re}\,\sigma_{\rm tf}$ is too small, when compared to 
experimental data.

The reduced normal current \eqref{lin6} contradicts microscopic 
studies \cite{Sch66,AT66,GE68,MRE93,HYY09}.
Indeed, the total current is derived from the Nambu-Gor'kov Green
function expanded in the gap, $G\approx G_0+G_0\Delta^*\tilde 
G_0\Delta G_0$, where $G_0$ gives ${\bf j}_n$ and the second
term provides the supercurrent. Apparently, the double-counting 
has to be  remedied  within the supercurrent itself.

With this issue in mind we shift to our new formulation 
of the theory, expressing the nucleation of superconductivity 
using the floating nucleation kernel. The Cooper pairs are 
created from electrons initially in the normal state, 
with mean velocity ${\bf v}={\bf J}_n/(en)$. The free 
energy of condensation has to supply the kinetic energy which 
electrons gain going from the normal component into the 
condensate, therefore the stability condition reads
\begin{align}
&{1\over 2m^*}\!\left(\!-i\hbar\nabla\!-\!
\frac{e^*}{c}{\bf A}\!-\!m^*{\bf v}\right)^2\!\varphi +\alpha\varphi+
\beta\!\left|\varphi\right|^2\!\varphi=-\!
\Gamma\partial_\tau\varphi.
\label{lin8}
\end{align}
We note that quantum kinetic energy is in fact a non-local 
contribution of the nucleation kernel. For the floating kernel
it depends exclusively on the  velocity differences  of the normal 
and superconducting component \cite{LL08}. 

The corresponding supercurrent
\begin{align}
\tilde{\bf j}_s={e^*\over m^*}{\rm Re}~
{\bar{\varphi}}\left(-i\hbar\nabla-
\frac{e^*}{c}{\bf A}-m^*{\bf v}\right)\varphi
\label{lin9}
\end{align}
we can write as $\tilde{\bf j}_s={\bf j}_s-e^*{\bf v}|\varphi|^2=
{\bf j}_s-(2|\varphi|^2/n){\bf J}_n$, therefore this approach is 
free of double-counting. 

If an effect of velocity $\bf v$ on the GL function is 
negligible, then $\varphi=\psi$ and the total current 
${\bf j}_{\rm fk}=
\tilde{\bf j}_s+{\bf J}_n$ obtained with the floating kernel 
is not different from the current in the two-fluid approximation 
${\bf j}_{\rm tf}={\bf j}_s+\tilde{\bf j}_n$. 
In the presence of vortices, the kinetic 
energy is non-zero due to diamagnetic currents and the 
perturbation enters the TDGL equation in the linear order 
leading to changes of the GL function. The averaged total 
current $\tilde {\bf J}_s+{\bf J}_n$ then differs from 
${\bf J}_s+\tilde {\bf J}_n$. The magneto-transmission thus
allows us to test the TDGL theory formulated with the 
floating nucleation kernel.

To obtain the conductivity we do not need to evaluate the 
modified GL function. The supercurrent modified by the
inertial force $m^*\partial_\tau{\bf v}$ is readily 
obtained from the condensate conductivity \eqref{lin3}. The 
driving force in Eq.~\eqref{lin9} is
$\partial_\tau\left(-(e^*/c){\bf A}-m^*{\bf v}\right)=
e^*{\bf E}+i(\omega/e^*n)\sigma_n{\bf E}$, therefore
\begin{align}
\tilde{\bf J}_s=\sigma_s\left(1+i\frac{\omega}{e^{*2}n}
\sigma_n\right){\bf E}.
\label{lin10}
\end{align}
The conductivity corresponding to the current
$\tilde{\bf J}_s+{\bf J}_n$ is given by
\begin{align}
\sigma_{\rm fk}=\sigma_s\left(1+i\frac{\omega}{e^{*2}n}
\sigma_n\right)+\sigma_n.
\label{lin11}
\end{align}
In Fig.~\ref{fig1} we compare ${\rm Im}\,\sigma_{\rm fk}$
with ${\rm Im}\,\sigma_s$. One can see that both values are
very close except for at the smallest magnetic field where 
${\rm Im}\,\sigma_{\rm fk}$ is closer to experimental data.

In contrast, the Joule heat obtained within various approximations 
is rather different. In Fig.~\ref{fig4} we compare the standard 
TDGL theory with the floating kernel formulation. Although none of the
approximations provides satisfactory values, among the tested 
approaches our floating kernel prescription leads to 
values closest to experiment.

In summary, we have formulated a version of TDGL theory using a floating 
nucleation kernel, meaning that the assumption of local equilibrium
is applied to electrons in the moving reference frame of the normal current.

When compared with standard TDGL theory in the context
of far-infrared spectroscopy, we have found that 
the floating kernel formulation yields better agreement with experiment.
In particular, recent published measurements of conductivity were considered;
since we have established the GL parameter $\kappa$ from 
the non-dissipative response given by the imaginary part of the
conductivity, our theory has no fitting parameters with respect 
to the Joule heat given by the real part of the conductivity.

Finally, since use of this new approach does not generally introduce
significant additional complexity, it may be promising
in the consideration of systems farther from equilibrium than is 
usually amenable to analysis via standard TDGL theory.

\medskip
The authors are grateful to Peter Matlock for valuable
comments and help in preparation of the manuscript. 
This work was supported by research plans
MSM 0021620834 and No. AVOZ10100521, by grants
GA\v{C}R 202/07/0597 and GAAV 100100712. 


\end{document}